\documentclass[sigconf, 10pt]{acmart}
\usepackage[acronyms,nonumberlist,nopostdot,nomain,nogroupskip]{glossaries}
\newacronym{quic}{QUIC}{Quick UDP Internet Connections}
\newacronym{3gpp}{3GPP}{3rd Generation Partnership Project}
\newacronym{adc}{ADC}{Analog to Digital Converter}
\newacronym{5g}{5G}{5th Generation}
\newacronym{aimd}{AIMD}{Additive Increase Multiplicative Decrease}
\newacronym{am}{AM}{Acknowledged Mode}
\newacronym{amc}{AMC}{Adaptive Modulation and Coding}
\newacronym{aqm}{AQM}{Active Queue Management}
\newacronym{awgn}{AGWN}{Additive White Gaussian Noise}
\newacronym{balia}{BALIA}{Balanced Link Adaptation}
\newacronym{bdp}{BDP}{Bandwidth-Delay Product}
\newacronym{bf}{BF}{Beamforming}
\newacronym{cc}{CC}{Congestion Control}
\newacronym{cdf}{CDF}{Cumulative Distribution Function}
\newacronym{ci}{CI}{Close-in free space reference}
\newacronym{cn}{CN}{Core Network}
\newacronym{cqi}{CQI}{Channel Quality Information}
\newacronym{cp}{CP}{Control Plane}
\newacronym{csirs}{CSI-RS}{Channel State Information - Reference Signal}
\newacronym{dc}{DC}{Dual Connectivity}
\newacronym{dce}{DCE}{Direct Code Execution}
\newacronym{dci}{DCI}{Downlink Control Information}
\newacronym{dl}{DL}{Downlink}
\newacronym{dmr}{DMR}{Deadline Miss Ratio}
\newacronym{dmrs}{DMRS}{DeModulation Reference Signal}
\newacronym{e2e}{E2E}{End-to-End}
\newacronym{ecn}{ECN}{Explicit Congestion Notification}
\newacronym{edf}{EDF}{Earliest Deadline First}
\newacronym{enb}{eNB}{evolved Node Base}
\newacronym{epc}{EPC}{Evolved Packet Core}
\newacronym{es}{ES}{Edge Server}
\newacronym{fdma}{FDMA}{Frequency Division Multiple Access}
\newacronym{fdd}{FDD}{Frequency Division Duplexing}
\newacronym[firstplural=Radio Access Technologies (RATs)]{rat}{RAT}{Radio Access Technology}
\newacronym{fs}{FS}{Fast Switching}
\newacronym{ftp}{FTP}{File Transfer Protocol}
\newacronym{gnb}{gNB}{Next Generation Node Base}
\newacronym{harq}{HARQ}{Hybrid Automatic Repeat reQuest}
\newacronym{hetnet}{HetNet}{Heterogeneous Network}
\newacronym{hh}{HH}{Hard Handover}
\newacronym{hol}{HOL}{Head-of-Line}
\newacronym{ia}{IA}{Initial Access}
\newacronym{imt}{IMT}{International Mobile Telecommunication}
\newacronym{iot}{IoT}{Internet of Things}
\newacronym{los}{LOS}{Line of Sight}
\newacronym{lte}{LTE}{Long Term Evolution}
\newacronym{m2m}{M2M}{Machine to Machine}
\newacronym{mac}{MAC}{Medium Access Control}
\newacronym{mc}{MC}{Multi-Connectivity}
\newacronym{mcs}{MCS}{Modulation and Coding Scheme}
\newacronym{mec}{MEC}{Mobile Edge Cloud}
\newacronym{mi}{MI}{Mutual Information}
\newacronym{mimo}{MIMO}{Multiple Input, Multiple Output}
\newacronym{mmwave}{mmWave}{millimeter wave}
\newacronym{mr}{MR}{Maximum Rate}
\newacronym{mss}{MSS}{Maximum Segment Size}
\newacronym{mtd}{MTD}{Machine-Type Device}
\newacronym{mtu}{MTU}{Maximum Transmission Unit}
\newacronym{nsf}{NSF}{National Science Foundation}
\newacronym{nfv}{NFV}{Network Function Virtualization}
\newacronym{nlos}{NLOS}{Non Line of Sight}
\newacronym{nr}{NR}{New Radio}
\newacronym{ofdm}{OFDM}{Orthogonal Frequency Division Multiplexing}
\newacronym{pdcch}{PDCCH}{Physical Downlonk Control Channel}
\newacronym{pdcp}{PDCP}{Packet Data Convergence Protocol}
\newacronym{pdsch}{PDSCH}{Physical Downlink Shared Channel}
\newacronym{pdu}{PDU}{Packet Data Unit}
\newacronym{pf}{PF}{Proportional Fair}
\newacronym{pgw}{PGW}{Packet Gateway}
\newacronym{phy}{PHY}{Physical}
\newacronym{pbch}{PBCH}{Physical Broadcast Channel}
\newacronym[plural=\gls{mme}s,firstplural=Mobility Management Entities (MMEs)]{mme}{MME}{Mobility Management Entity}
\newacronym{prb}{PRB}{Physical Resource Block}
\newacronym{pss}{PSS}{Primary Synchronization Signal}
\newacronym{pucch}{PUCCH}{Physical Uplink Control Channel}
\newacronym{pusch}{PUSCH}{Physical Uplink Shared Channel}
\newacronym{rach}{RACH}{Random Access Channel}
\newacronym{ran}{RAN}{Radio Access Network}
\newacronym{red}{RED}{Random Early Detection}
\newacronym{rf}{RF}{Radio Frequency}
\newacronym{rlc}{RLC}{Radio Link Control}
\newacronym{rlf}{RLF}{Radio Link Failure}
\newacronym{rrc}{RRC}{Radio Resource Control}
\newacronym{rrm}{RRM}{Radio Resource Management}
\newacronym{rr}{RR}{Round Robin}
\newacronym{rs}{RS}{Remote Server}
\newacronym{rsrp}{RSRP}{Reference Signal Received Power}
\newacronym{rss}{RSS}{Received Signal Strength}
\newacronym{rtt}{RTT}{Round Trip Time}
\newacronym{rw}{RW}{Receive Window}
\newacronym{rx}{RX}{Receiver}
\newacronym{sa}{SA}{standalone}
\newacronym{sack}{SACK}{Selective Acknowledgment}
\newacronym{sap}{SAP}{Service Access Point}
\newacronym{sch}{SCH}{Secondary Cell Handover}
\newacronym{scoot}{SCOOT}{Split Cycle Offset Optimization Technique}
\newacronym{sdma}{SDMA}{Spatial Division Multiple Access}
\newacronym{sinr}{SINR}{Signal to Interference plus Noise Ratio}
\newacronym{sm}{SM}{Saturation Mode}
\newacronym{snr}{SNR}{Signal to Noise Ratio}
\newacronym{son}{SON}{Self-Organizing Network}
\newacronym{ss}{SS}{Synchronization Signal}
\newacronym{srs}{SRS}{Sounding Reference Signal}
\newacronym{sss}{SSS}{Secondary Synchronization Signal}
\newacronym{tb}{TB}{Transport Block}
\newacronym{tcp}{TCP}{Transmission Control Protocol}
\newacronym{tdd}{TDD}{Time Division Duplexing}
\newacronym{tdma}{TDMA}{Time Division Multiple Access}
\newacronym{tfl}{TfL}{Transport for London}
\newacronym{thz}{THz}{Terahertz}
\newacronym{tm}{TM}{Transparent Mode}
\newacronym{trp}{TRP}{Transmitter Receiver Pair}
\newacronym{tti}{TTI}{Transmission Time Interval}
\newacronym{ttt}{TTT}{Time-to-Trigger}
\newacronym{tx}{TX}{Transmitter}
\newacronym{ue}{UE}{User Equipment}
\newacronym{ul}{UL}{Uplink}
\newacronym{uml}{UML}{Unified Modeling Language}
\newacronym{um}{UM}{Unacknowledged Mode}
\newacronym{utc}{UTC}{Urban Traffic Control}
\newacronym{vm}{VM}{Virtual Machine}
\newacronym{rsrq}{RSRQ}{Reference Signal Received Quality}
\newacronym{rssi}{RSSI}{Received Signal Strength Indicator}
\newacronym{crs}{CRS}{Cell Reference Signal}
\newacronym{comp}{CoMP}{Coordinated Multi-Point}
\newacronym{cran}{C-RAN}{Cloud \acrlong{ran}}
\newacronym{ca}{CA}{Carrier Aggregation}
\newacronym{cco}{CC}{Carrier Component}
\newacronym{nsa}{NSA}{Non Stand Alone}
\newacronym{embb}{eMBB}{Enhanced Mobility Broadband}
\newacronym{bsr}{BSR}{Buffer Status Report}
\newacronym{srb}{SRB}{Service Radio Bearer}
\newacronym{scm}{SCM}{Spatial Channel Model}
\newacronym{sctp}{SCTP}{Stream Control Transmission Protocol}
\newacronym{mptcp}{MPTCP}{Multi-path TCP}
\newacronym{ietf}{IETF}{Internet Engineering Task Force}
\newacronym{os}{OS}{Operating System}
\newacronym{tls}{TLS}{Transport Layer Security}
\newacronym{rfc}{RFC}{Request for Comments}
\newacronym{http}{HTTP}{HyperText Transfer Protocol}
\newacronym{nat}{NAT}{Network Address Translation}
\newacronym{api}{API}{Application Programming Interface}
\newacronym{rto}{RTO}{Retransmission Timeout}
\newacronym{psc}{PSC}{Public Safety Communication}
\newacronym{rpgm}{RPGM}{Reference Point Group Mobility}
\newacronym{ic}{IC}{Incident Command}
\newacronym{rsu}{RSU}{Road Side Unit}
\newacronym{uav}{UAV}{Unmanned Aerial Vehicle}
\newacronym{usa}{U.S.}{United States}
\newacronym{vr}{VR}{Virtual Reality}
\newacronym{iab}{IAB}{Integrated Access and Backhaul}
\newacronym{wlan}{WLAN}{Wireless Local Area Network}
\newacronym{cots}{COTS}{Commercial Off-the-Shelf}
\newacronym{fpga}{FPGA}{Field Programmable Gate Array}
\newacronym{rcn}{RCN}{Research Coordination Network}
\newacronym{abg}{ABG}{Alpha-Beta-Gamma}
\newacronym{fi}{FI}{Floating Intercept}
\newacronym{uas}{UAS}{Unmanned Aerial System}
\newacronym{gps}{GPS}{Global Positioning System}
\newacronym{a2g}{A2G}{air-to-ground}
\newacronym{a2a}{A2A}{air-to-air}
\newacronym{uma}{UMa}{Urban Macro}
\newacronym{umi}{UMi}{Urban Micro}
\newacronym{rma}{RMa}{Rural Macro}
\newacronym{inoo}{InOo}{Indoor Open Office}
\newacronym{ple}{PLE}{path loss exponent}

\usepackage{tikz}
\usepackage{pgfplots}
\pgfplotsset{compat=newest} 
\pgfplotsset{plot coordinates/math parser=false} 
\newlength\fheight
\newlength\fwidth
\usetikzlibrary{plotmarks,patterns,decorations.pathreplacing,backgrounds,calc,arrows,arrows.meta,spy,matrix,decorations.markings}
\usepgfplotslibrary{patchplots,groupplots}
\usepackage{tikzscale}

\tikzstyle{startstop} = [rectangle, rounded corners, minimum width=2cm, minimum height=0.5cm,text centered, draw=black]
\tikzstyle{io} = [trapezium, trapezium left angle=70, trapezium right angle=110, minimum width=3cm, minimum height=1cm, text centered, draw=black]
\tikzstyle{process} = [rectangle, minimum width=2cm, minimum height=0.5cm, text centered, draw=black, align=center]
\tikzstyle{decision} = [ellipse, minimum width=2cm, minimum height=1cm, text centered, draw=black]
\tikzstyle{arrow} = [thick,<->,>=stealth]
\tikzstyle{line} = [thick,>=stealth]
\tikzstyle{lineDashed} = [thick,>=stealth,dashed]
\tikzstyle{darrow} = [thick,<->,>=stealth]
\tikzstyle{sarrow} = [thick,->,>=stealth]
\tikzstyle{larrow} = [line width=0.1mm,dashdotted,<->,>=stealth]
\tikzstyle{vecArrow} = [thick, decoration={markings,mark=at position
   1 with {\arrow[semithick, fill=white]{open triangle 60}}},
   double distance=1.4pt, shorten >= 5.5pt,
   preaction = {decorate},
   postaction = {draw,line width=1.4pt, white,shorten >= 4.5pt}]
\tikzstyle{innerWhite} = [semithick, white,line width=1.4pt, shorten >= 4.5pt]

\definecolor{SchoolColor}{RGB}{0.71, 0, 0.106}
\definecolor{chaptergrey}{rgb}{0.61, 0, 0.09} 
\definecolor{midgrey}{rgb}{0.4, 0.4, 0.4}
\definecolor{chaptergreen}{rgb}{0.09, 0.612, 0}
\definecolor{chapterpurple}{rgb}{0.522, 0, 0.612}
\definecolor{chapterlightgreen}{rgb}{0, 0.612, 0.522}

\makeatletter
\def\grd@save@target#1{%
  \def\grd@target{#1}}
\def\grd@save@start#1{%
  \def\grd@start{#1}}
\tikzset{
  grid with coordinates/.style={
    to path={%
      \pgfextra{%
        \edef\grd@@target{(\tikztotarget)}%
        \tikz@scan@one@point\grd@save@target\grd@@target\relax
        \edef\grd@@start{(\tikztostart)}%
        \tikz@scan@one@point\grd@save@start\grd@@start\relax
        \draw[minor help lines] (\tikztostart) grid (\tikztotarget);
        \draw[major help lines] (\tikztostart) grid (\tikztotarget);
        \grd@start
        \pgfmathsetmacro{\grd@xa}{\the\pgf@x/1cm}
        \pgfmathsetmacro{\grd@ya}{\the\pgf@y/1cm}
        \grd@target
        \pgfmathsetmacro{\grd@xb}{\the\pgf@x/1cm}
        \pgfmathsetmacro{\grd@yb}{\the\pgf@y/1cm}
        \pgfmathsetmacro{\grd@xc}{\grd@xa + \pgfkeysvalueof{/tikz/grid with coordinates/major step x}}
        \pgfmathsetmacro{\grd@yc}{\grd@ya + \pgfkeysvalueof{/tikz/grid with coordinates/major step y}}
        \foreach \x in {\grd@xa,\grd@xc,...,\grd@xb}
        \node[anchor=north] at (\x,\grd@ya) {\pgfmathprintnumber{\x}};
        \foreach \y in {\grd@ya,\grd@yc,...,\grd@yb}
        \node[anchor=east] at (\grd@xa,\y) {\pgfmathprintnumber{\y}};
      }
    }
  },
  minor help lines/.style={
    help lines,
    gray,
    line cap =round,
    xstep=\pgfkeysvalueof{/tikz/grid with coordinates/minor step x},
    ystep=\pgfkeysvalueof{/tikz/grid with coordinates/minor step y}
  },
  major help lines/.style={
    help lines,
    line cap =round,
    line width=\pgfkeysvalueof{/tikz/grid with coordinates/major line width},
    xstep=\pgfkeysvalueof{/tikz/grid with coordinates/major step x},
    ystep=\pgfkeysvalueof{/tikz/grid with coordinates/major step y}
  },
  grid with coordinates/.cd,
  minor step x/.initial=.5,
  minor step y/.initial=.2,
  major step x/.initial=1,
  major step y/.initial=1,
  major line width/.initial=1pt,
}
\makeatother
\usepackage[font=small]{caption}
\usepackage[font=small]{subcaption}
\usepackage{amsmath}
\usepackage{gensymb}

\usepackage{wrapfig}
\usepackage[keeplastbox]{flushend}
\usepackage{booktabs}
\usepackage[inline]{enumitem}
\usepackage{xspace}

\newcommand{\freqGhz}[1]{$#1\:\mathrm{GHz}$}
\newcommand{\meter}[1]{$#1\:\mathrm{m}$}

\usepackage{dblfloatfix}    

\newlist{romaninline}{enumerate*}{1}
\setlist[romaninline]{label=(\roman*)}

\newcommand{\dji}[0]{DJI M$600$\xspace}
\newcommand{\terragraph}[0]{Facebook Terragraph\xspace}
\newcommand{\nuc}[0]{Intel NUC\xspace}

\title{An Experimental mmWave Channel Model\\for UAV-to-UAV Communications}

\author{Michele Polese, Lorenzo Bertizzolo, Leonardo Bonati}
\author{Abhimanyu Gosain, Tommaso Melodia}

\affiliation{
\institution{Institute for the Wireless Internet of Things, Northeastern University, Boston, MA 02115, USA\\
Email: \{m.polese, bertizzolo.l, bonati.l, a.gosain, t.melodia\}@northeastern.edu
}
}

\begin{abstract}
    \gls{uav} networks can provide a resilient communication infrastructure to enhance terrestrial networks in case of traffic spikes or disaster scenarios. However, to be able to do so, they need to be based on high-bandwidth wireless technologies for both radio access and backhaul.
    With this respect, the \gls{mmwave} spectrum represents an enticing  solution, since it provides large chunks of untapped spectrum that can enable ultra-high data-rates for aerial platforms. 
    Aerial \gls{mmwave} channels, however, experience characteristics that are significantly different from terrestrial deployments in the same frequency bands. As of today, mmWave aerial channels have not been extensively studied and modeled. Specifically, the combination of \gls{uav} micro-mobility (because of imprecisions in the control loop, and external factors including wind) and the highly directional \gls{mmwave} transmissions require ad hoc models to accurately capture the performance of \gls{uav} deployments. To fill this gap, we propose an empirical propagation loss model for \gls{uav}-to-\gls{uav} communications at \freqGhz{60}, based on an extensive aerial measurement campaign conducted with the Facebook Terragraph channel sounders. We compare it with 3GPP channel models and make the measurement dataset publicly available. 
\end{abstract}


\setcopyright{acmlicensed}
\begin{document}

\fancyhead{}

\copyrightyear{2020} 
\acmYear{2020} 
\acmConference[mmNets'20]{4th ACM Workshop on Millimeter-wave Networks and Sensing Systems}{September 25, 2020}{London, UK}
\acmBooktitle{4th ACM Workshop on Millimeter-wave Networks and Sensing Systems (mmNets'20), September 25, 2020, London, UK}
\acmPrice{15.00}
\acmDOI{10.1145/xxxxxxx.xxxxxxxx}
\acmISBN{xxxxxxxxxxxxxxxxxxxxxxxxx}
\maketitle

\begin{picture}(300,0)(300,-240)
\put(0,0){
\put(0,0){This paper has been accepted at 4th ACM Workshop on Millimeter-wave Networks and Sensing Systems (mmNets'20).}
\put(0,-15){Copyright may be transferred without notice.}
}
\end{picture}

\glsresetall

\section{Introduction}
\label{sec:intro}
Unmanned aerial systems are promising technological enablers for the wireless industry, as they provide an effective and inexpensive solution to temporarily connect ground users in the absence of terrestrial infrastructure \cite{BertizzoloInfocom20SwarmControl, BertizzoloHotmobile20, BertizzoloMmnets19}. 
In this domain, a key research challenge is how to provide high-capacity robust backhaul and inter-\gls{uav} connectivity in flying platforms. 
Fiber optic backhaul, typical of terrestrial infrastructure, is not a feasible solution for \glspl{uav}, thus, both the radio access and the backhaul exploit wireless links.

However, combining access and backhaul on the same wireless interface introduces tight data-rate and latency requirements to the underlying communication technology.
This problem is exacerbated when multiple \glspl{uav} rely on each other for data forwarding, in an aerial multi-hop fashion.
While traditional sub-\freqGhz{6} technologies are unfit for high-load traffic aggregation, 
the \gls{mmwave} spectrum can be a unified solution for fully-wireless nodes offering unprecedented bandwidth.
Thus, at \glspl{mmwave}, a \gls{uav} network can use the same wireless technology to provide connectivity to ground users, communicate with neighboring \glspl{uav}, and relay data traffic toward the closest ground tower, in an ``integrated access and backhaul'' fashion \cite{polese2020integrated}.

Even though recent 5G standard specifications already envision the use of \glspl{mmwave} \cite{3gpp.38.300}, and some previous work studied their propagation for ground deployments \cite{rappaport2017overview,3gpp.38.901}, the aerial wireless channel at \gls{mmwave} frequencies has not been extensively characterized yet. 
Indeed, \glspl{mmwave} may affect the communication quality in aerial scenarios differently from ground deployments, for the following reasons:
\begin{romaninline}
\item Some frequencies in the \gls{mmwave} spectrum (e.g., the 60 GHz band) suffer from oxygen absorption. As these atmospheric conditions change with the deployment height, high-altitude aerial scenarios might differ from ground deployments; 
\item even when \gls{gps}-locked, the inaccuracy of \gls{uav}'s on-board sensors may lead to slight horizontal (yaw) and vertical (throttle) drone fluctuations. Given that \glspl{mmwave} exploit highly directional communications, these fluctuations might severely deteriorate the channel quality~\cite{BertizzoloMmnets19}; 
\item when \glspl{uav} fly in harsh wind conditions, they lean forward/backward (pitch) and sideways (roll) to counterbalance the wind force. Tilting a \gls{uav}-mounted \gls{mmwave} radio might compromise the link quality by changing the best beam path or the radios' polarization~\cite{KhuwajaSurvey}; 
\item on-board \gls{mmwave} radios are often mounted either above or below the main \gls{uav}'s frame structure. The frame size and its material, together with the spinning propellers introduce ambient noise right where the headset is mounted. On-board batteries, radio control, and circuitry non-linearity exacerbate this effect known as ``airframe shadowing''~\cite{semkin2020rcs}.
\end{romaninline}
For these reasons, the performance evaluation of \gls{mmwave} \gls{uav}-to-\gls{uav} communications cannot be based upon ground-tailored \gls{mmwave} channel studies and demands the development of dedicated \gls{a2a} propagation and fading models. Prior work has focused on analytical or ray-tracing approaches~\cite{khawaja2017mmWaveUAVChannel, khawaja2018temporal, GapeyenkoJSAC, DabiriTWC}, which have not been validated through experimental measurement campaigns.




In this paper, we propose an empirical propagation loss model for \gls{a2a} communications at \freqGhz{60}. It is based on an extensive measurement campaign, with more than 3 days of flight experiments, and the \terragraph channel sounders~\cite{terragraph} mounted on two \glspl{uav}. 
The measurements validate the empirical model in a wide range of flying heights (\meter{6-15}) and distances (\meter{6-40}), and show that, in the considered range, the path loss does not have an explicit dependence on the \gls{uav} height. Moreover, we compare the path loss curve with 3GPP channel models, and, using the same measurement campaign, we characterize the impact of a sub-optimal beam selection on the link budget. \textit{To the best of our knowledge, this is the first empirical \gls{a2a} propagation model for the 60 GHz band modeling the impact of the \glspl{uav} micro-mobility on the channel.} 
Last, we publicly release the collected measurements' dataset and analysis scripts to the community.\footnote{\url{https://github.com/wineslab/uav-to-uav-60-ghz-channel-model}} 

The remainder of the paper is organized as follows. Sec.~\ref{sec:meas} describes the measurement campaign. Sec.~\ref{sec:data} analyzes empirical path loss fits on the collected data and compares them with terrestrial channel models. Finally, Sec.~\ref{sec:conclusions} concludes the paper.


\begin{figure}[b]
    \begin{subfigure}[t]{0.48\textwidth}
        \centering
        \setlength\belowcaptionskip{0cm}
        \includegraphics[width=0.65\columnwidth]{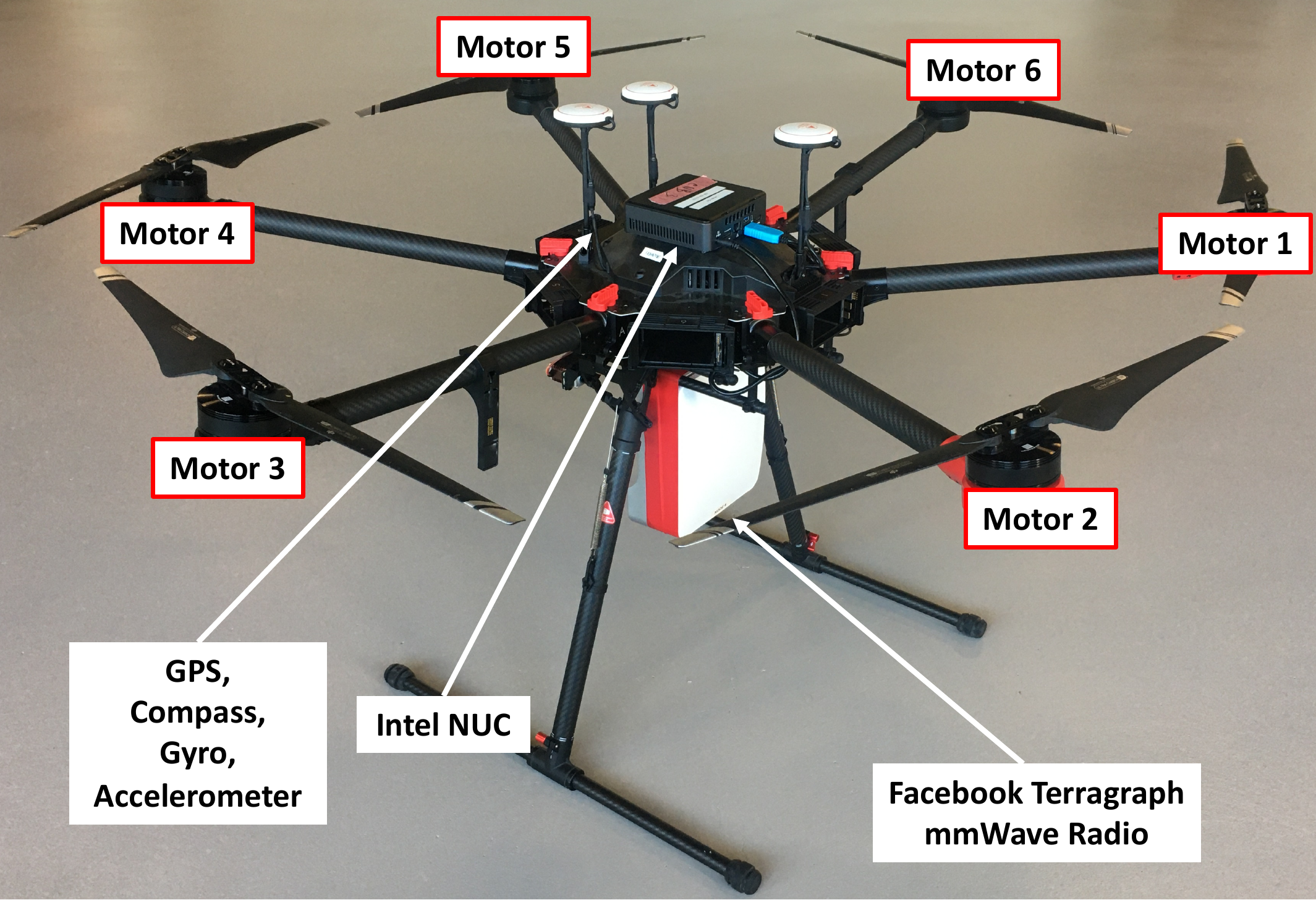}
        \vspace{-2mm}
    \end{subfigure}
    \begin{subfigure}[t]{0.48\textwidth}
        \centering
        \setlength\belowcaptionskip{0cm}
        \includegraphics[width=1\columnwidth]{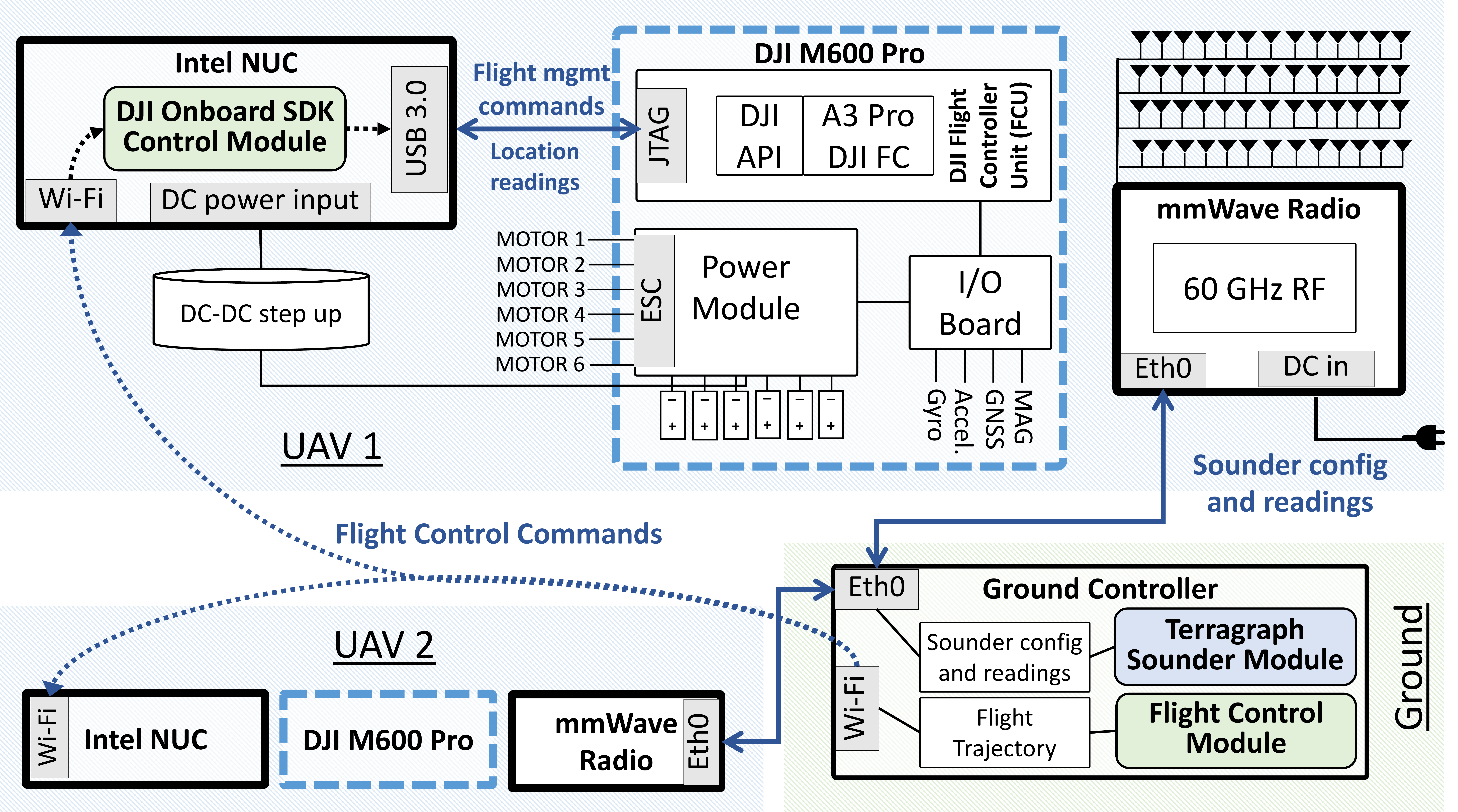}
    \end{subfigure}
    \setlength\belowcaptionskip{0pt}
    \caption{\gls{uav}-mounted \gls{mmwave} channel sounder: Prototype and hardware schematics.}
    \label{fig:prototype_complete}
\end{figure}

\section{Measurement Campaign Setup}
\label{sec:meas}

For our measurements, we employed two \dji as \glspl{uav} and two \terragraph \gls{mmwave} radios configured as channel sounders with beam scanning capabilities.
Each \dji mounts an on-board powered \nuc computer to perform flight control tasks, while the \terragraph radios are powered from the ground and coordinated in their channel sounding procedures by a ground host controller. The hardware and software schematics of the \gls{mmwave}-enabled \gls{uav} are reported in Fig.~\ref{fig:prototype_complete}.

The \terragraph channel sounders operate in the IEEE 802.11ad bands~\cite{terragraphindoor}. 
The radios feature TX and RX arrays of $36\:\times\:8$ antenna elements with an angular coverage of $90^{\circ}$ and $64$ beam directions in the azimuth plane, thus with a spacing between each beam of $1.4^{\circ}$. The half-power beamwidth is $2.8^{\circ}$, and the radio maximum effective radiated power is $45\:\mathrm{dBm}$. Moreover, the antenna, circuitry, and main board are enclosed in a rugged case, with a small weight and form factor that makes it possible to deploy them on \glspl{uav}.
\terragraph sounders have recently been used for channel measurement campaigns in the \freqGhz{60} band by several research groups. They are calibrated following the procedure in~\cite{terragraphindoor} and allow received power, path loss, and \gls{snr} measurements for different transmit and receive beam pairs. The standard deviation in path loss measurements within 1 dB~\cite{terragraphindoor}. 
Some of \terragraph previous use and documentation can be found at \cite{BertizzoloMmnets19, rasekh2017noncoherent, terragraphsara, terragraphindoor, terragraphnlos}.

\begin{figure}[t]
\setlength\abovecaptionskip{2pt}
        \centering
        \includegraphics[width=\columnwidth]{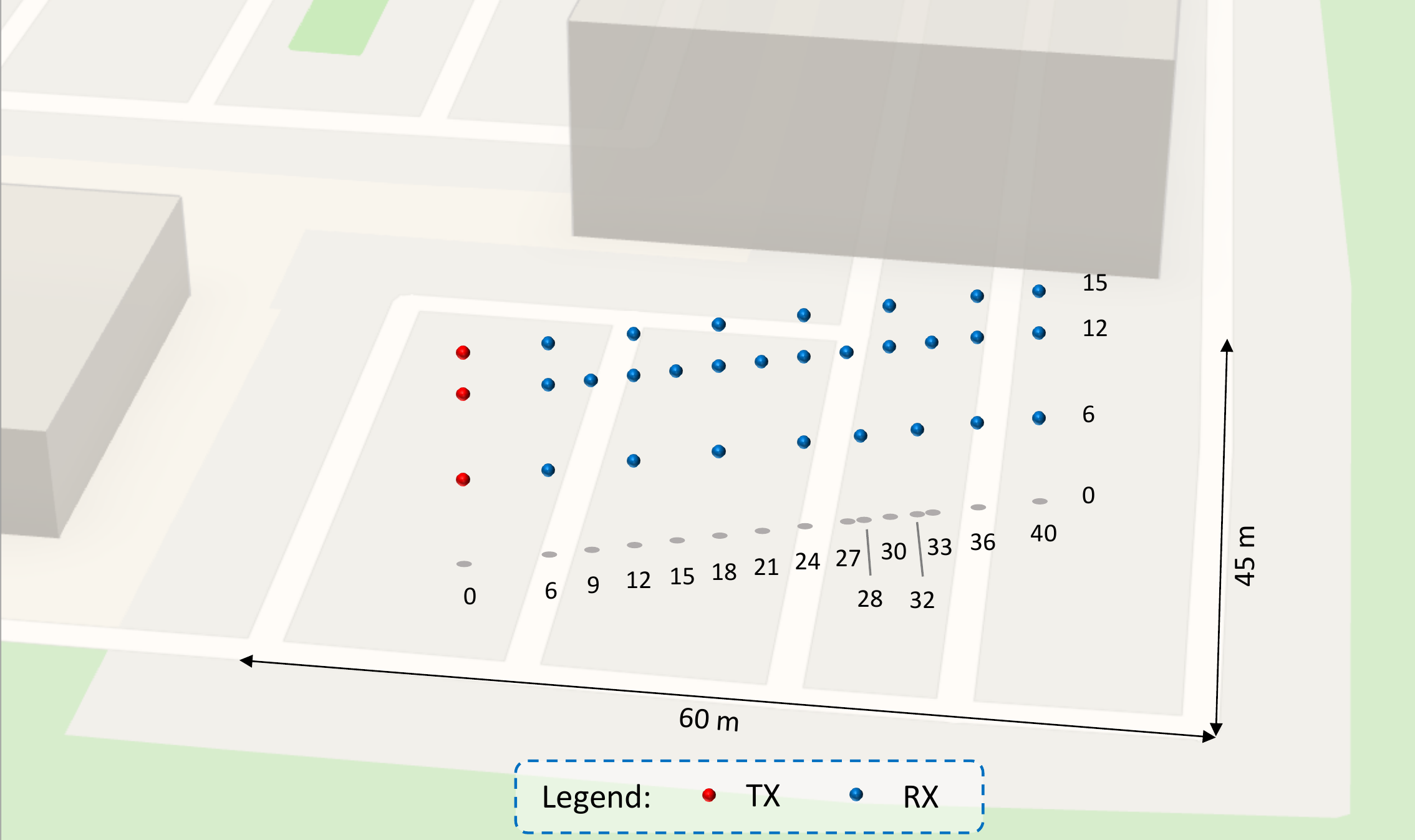}
        \caption{Deployment scenario.}
    \label{fig:deployment}
\end{figure}

In our measurement campaign, we run extensive channel sounding experiments on a wide open field with few reflectors and scarce multi-path effect as illustrated in Fig.~\ref{fig:deployment}.\footnote{A video of the experiments is available at \url{https://youtu.be/Jzwt-tEp98g}.} For each experiment, the transmitter and receiver \glspl{uav} hover at the same altitude and face each other in full \gls{los} conditions.
Through our experiment, we consider $3$ heights, \meter{6}, \meter{12}, and \meter{15}; and a total of $14$ distances between transmitter and receiver, namely 6, 9, 12, 15, 18, 21, 24, 27, 28, 30, 32, 33, 36, and \meter{40}, as shown in Fig.~\ref{fig:deployment}. The selection of these coordinates has been constrained by the size of the \gls{uav} flying facility.
We performed channel measurements for channel~$2$ of the IEEE 802.11ad standard (i.e., the carrier frequency is \freqGhz{60.48}, and the bandwidth if \freqGhz{2.16}) and $27$ altitude-distance pairs, accounting for 3 days of flight experiments.
Each channel sounding experiment consists of scanning the $400$ beam pairs, between $\pm 14^{\circ}$ from the boresight direction (see \cite{terragraphindoor} for specifications), with $15$ independent measurements per beam scan, to average small scale fading, and a distance-adaptive transmission gain.

\section{Path Loss Analysis}
\label{sec:data}


This section describes the data analysis based on the measurement campaign outlined in Sec.~\ref{sec:meas}.

\subsection{Experimental Path Loss Laws}

The literature on propagation models has proposed several experimental path loss laws that can be used to fit the measurement results as a function of distance and frequency parameters. A review can be found in~\cite{rappaport2017overview}. In this paper, as the measurements have been taken for a single carrier frequency (i.e., 60.48 GHz), we focus on a distance-dependent fit only.

The \gls{ci} path loss models can be expressed as~\cite{rappaport2017overview}
\begin{equation}\label{eq:ci}
	PL_{CI}(d, f) = PL_{{\rm FS}, ref}(f) + 10 n_{CI} \log_{10}(d) + \xi_{\sigma, CI}.
\end{equation}
The first term, $PL_{{\rm FS}, ref}(f)$, is used to model the dependence on the carrier frequency, and is calculated using Friis' law for free space propagation, at the reference distance of 1 m~\cite{sun2015pathloss}:
\begin{equation}
	PL_{{\rm FS}, ref}(f) = 20 \log_{10} \left(\frac{4 \pi f}{c}\right),
\end{equation}
where $c$ is the speed of light. The second term accounts for the logarithmic distance-dependent behavior, with $n_{CI}$ the \gls{ple}, given by the value that best fits the measurement data. Finally, $\xi_{\sigma, CI}$ is a shadow fading term that in the decibel domain follows a Gaussian distribution with zero mean and standard deviation $\sigma$. The \gls{ci} model is widely used for empirical path loss fitting, either in the single-slope version of Eq.~\eqref{eq:ci}~\cite{rappaport2017overview,3gpp.38.901}, or with a dual-slope extension, in which different values of $n_{CI}$ are considered before and after a breakpoint distance~\cite{3gpp.38.901,ghosh20155g}. 

Other widespread models belong to the \gls{fi} family, in which the term based on Friis' law is replaced by a generic value, determined based on the best fit on the data. A notable example is the \gls{abg} model, given by~\cite{rappaport2017overview}
\begin{equation}\label{eq:abg_full}
	PL_{ABG}(d,f) = \beta + 10 \gamma \log_{10}(f) + 10 \alpha \log_{10} (d) + \xi_{\sigma, ABG},
\end{equation}
where $\xi_{\sigma, ABG}$ is a Gaussian shadow fading with zero mean and standard deviation $\sigma$, as in the \gls{ci} model,	and $\beta$, $\gamma$, and $\alpha$ are fit on the data. With respect to the \gls{ci} model in Eq.~\eqref{eq:ci}, the term $\alpha$ is equivalent to the path loss exponent $n_{CI}$. Notice that, as the measurement campaign of this paper is based on a single frequency, we cannot compute both $\beta$ and $\gamma$. Therefore, in the following, we will consider a simplified version, with two fit parameters, given by
\begin{equation}\label{eq:fi}
	PL_{FI}(d) = PL_{FI} + 10 n_{FI} \log_{10}(d) + \xi_{\sigma, FI},
\end{equation}
where $PL_{FI} = \beta + 10 \gamma \log_{10}(f)$, $n_{FI} = \alpha$, and $\xi_{\sigma, FI} = \xi_{\sigma, ABG}$.

For both models, the fit parameters (i.e., $n_{CI}$ for \gls{ci}, and $n_{FI}$ and $PL_{FI}$ for \gls{fi}) are computed as the slope (and the intercept, for \gls{fi}) of a linear fit on the logarithm of the distance, and the standard deviation of the shadow fading is computed as the root mean square error on the fit~\cite{rappaport2017overview,amorim2017radio}. 


\subsection{Comparison of CI and FI Fits}

\begin{figure}[t]
    \centering
    \setlength\fheight{.4\columnwidth}
    \setlength\fwidth{.9\columnwidth}
%
%
\definecolor{mycolor1}{rgb}{0.00000,0.44700,0.74100}%
\definecolor{mycolor2}{rgb}{0.85000,0.32500,0.09800}%
\definecolor{mycolor3}{rgb}{0.92900,0.69400,0.12500}%
\definecolor{mycolor4}{rgb}{0.49400,0.18400,0.55600}%
\definecolor{mycolor5}{rgb}{0.46600,0.67400,0.18800}%
\definecolor{mycolor6}{rgb}{0.30100,0.74500,0.93300}%
\begin{tikzpicture}
\pgfplotsset{every tick label/.append style={font=\scriptsize}}

\begin{axis}[%
width=0.951\fwidth,
height=\fheight,
at={(0\fwidth,0\fheight)},
scale only axis,
xmin=5,
xmax=45,
xlabel style={font=\footnotesize\color{white!15!black}},
xlabel={UAV-to-UAV distance [m]},
ymin=80,
ymax=110,
ylabel style={font=\footnotesize\color{white!15!black}},
ylabel={Pathloss [dB]},
axis background/.style={fill=white},
axis x line*=bottom,
axis y line*=left,
xmajorgrids,
ymajorgrids,
legend style={legend cell align=left, align=left, draw=white!15!black, font=\scriptsize, at={(0.5,0.95)}, anchor=south},
legend columns=3
]

\addplot[only marks, mark=asterisk, mark options={}, mark size=1.5000pt, draw=mycolor5] table[row sep=crcr]{%
x	y\\
6	83.1280544048846\\
12	88.9876700318486\\
18	93.4918541481491\\
24	102.287628326341\\
30	100.058029873338\\
36	105.505665325063\\
40	107.934794437175\\
};
\addlegendentry{Meas., $h = 6$ m}

\addplot[only marks, mark=+, mark options={}, mark size=1.5000pt, draw=mycolor3] table[row sep=crcr]{%
x	y\\
6	86.0397539475613\\
9	91.5460418808438\\
12	94.4062366913422\\
15	97.9879394678565\\
18	95.7170813033622\\
21	95.3777219945156\\
24	99.02263962\\
27	99.2763817738388\\
30	102.051493191169\\
33	100.437992682641\\
36	102.844470979747\\
40	103.9426507\\
};
\addlegendentry{Meas., $h = 12$ m}

\addplot[only marks, mark=o, mark options={}, mark size=1.5000pt, draw=mycolor4] table[row sep=crcr]{%
x	y\\
6	85.2846006761252\\
12	91.9142447575216\\
18	97.7809256435052\\
24	97.5739435439125\\
28	100.3026762\\
32	101.3044984\\
36	101.9418107\\
40	104.8675908\\
};
\addlegendentry{Meas., $h = 15$ m}

\addplot [color=mycolor1]
  table[row sep=crcr]{%
6	85.5996542949231\\
9	89.5642497031066\\
12	92.3771749108529\\
15	94.5590492107989\\
18	96.3417703190363\\
21	97.8490395394706\\
24	99.1546955267826\\
27	100.30636572722\\
30	101.336569826729\\
33	102.268502793707\\
36	103.119290934966\\
39	103.9019401567\\
42	104.6265601554\\
};
\addlegendentry{CI fit, all heights}

\addplot [color=mycolor2, dashdotted]
  table[row sep=crcr]{%
6	85.1503061774636\\
9	89.2516809356505\\
12	92.1616526366479\\
15	94.4188020448483\\
18	96.2630273948348\\
21	97.8222977107215\\
24	99.1729990958322\\
27	100.364402153022\\
30	101.430148504033\\
33	102.394233348402\\
36	103.274373854019\\
39	104.084024633493\\
42	104.833644169906\\
};
\addlegendentry{FI fit, all heights}

\end{axis}
\end{tikzpicture}%
        \setlength\abovecaptionskip{-.1cm}
    \caption{Comparison of CI and FI fits.}
    \label{fig:ci_abg}
\end{figure}
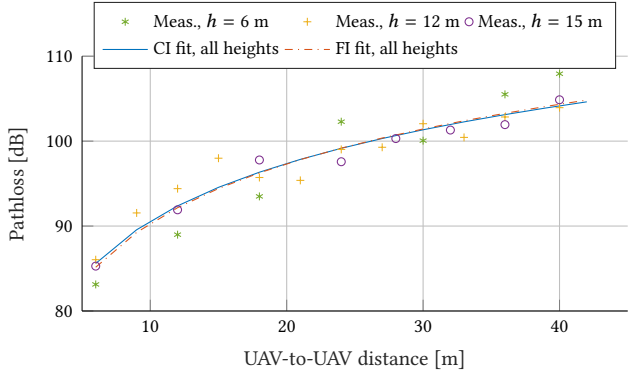

We first discuss whether a \gls{ci} fit is representative of the \gls{a2a} path loss measurements, or whether a \gls{fi} fit is preferable. Figure~\ref{fig:ci_abg} compares the two methods, considering data for all the values for the height (i.e., $h \in [6, 12, 15]$ m) and distance $d$ from $6$ to \meter{40}. Both curves share the same trend, with the \gls{fi} fit slightly steeper, but with a difference of at most $0.5\:\mathrm{dB}$ at $d=\:$\meter{6} and $0.2\:\mathrm{dB}$ at $d=\:$\meter{40}. Table~\ref{fig:ci_abg} reports the fit parameters for both, showing that the \gls{fi} identifies a fit for its floating intercept at a $67.03\:\mathrm{dB}$, which is $1.05\:\mathrm{dB}$ smaller than the free space path loss in the same conditions, and a path loss exponent of $2.33$, against the value of $2.25$ for the \gls{ci} fit. 
The shadowing standard deviation $\sigma$ is $0.04\:\mathrm{dB}$ smaller for the \gls{fi} fit, which is almost two orders of magnitude smaller than the actual value of $\sigma$ (i.e., $3.52\:\mathrm{dB}$ for \gls{fi} and $3.56\:\mathrm{dB}$ for \gls{ci}). Following standard practices in the literature~\cite{sun2015pathloss}, we conclude that both models provide a suitable representation of the path loss in a \gls{a2a} \freqGhz{60} link. Given this, in the remainder of the paper, we will consider the \gls{ci} fit as baseline, as it is simpler than the \gls{fi} fit (i.e., one fit parameter instead of two), and is based on the fundamental principles of wireless propagation through the Friis-based intercept~\cite{sun2015pathloss}.

\begin{table}
    \centering
    \footnotesize
    \begin{tabular}{llll}
    \toprule
        & Intercept [dB] & Path loss exponent & $\sigma$ [dB] \\\midrule
    CI fit &  68.08  & 2.25 & 3.56 \\
    FI fit & 67.03 & 2.33 & 3.52\\
    \bottomrule
    \end{tabular}
    \setlength\abovecaptionskip{-.1cm}
    \caption{Comparison of the parameters for CI and FI fits, considering all heights $h\in[6,12,15]$~m.}
    \label{tab:ci_abg}
\end{table}

\subsection{Impact of the Height on Path Loss}

\begin{table}[b]
    \centering
    \footnotesize
    \begin{tabular}{lllll}
    \toprule
        & \bf All heights & $h=6$ m & $h=12$ m & $h=15$ m\\\midrule
    PLE $n_{CI}$ & \bf 2.25 & 2.23 & 2.25 & 2.28\\
    $\sigma$ [dB] & \bf 3.56 & 0.82 & 2.62 & 8.06\\
    \bottomrule
    \end{tabular}
    \caption{Parameters for the CI fit for different heights.}
    \label{tab:ci_height}
\end{table}

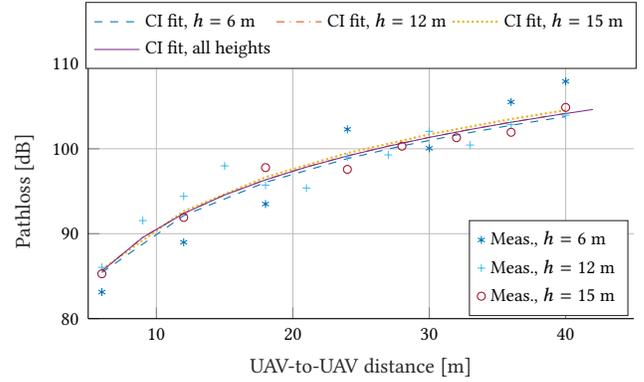
\begin{figure}[t]
    \centering
    \setlength\fheight{.4\columnwidth}
    \setlength\fwidth{.9\columnwidth}
%
%
\definecolor{mycolor1}{rgb}{0.00000,0.44700,0.74100}%
\definecolor{mycolor2}{rgb}{0.85000,0.32500,0.09800}%
\definecolor{mycolor3}{rgb}{0.92900,0.69400,0.12500}%
\definecolor{mycolor4}{rgb}{0.49400,0.18400,0.55600}%
\definecolor{mycolor5}{rgb}{0.46600,0.67400,0.18800}%
\definecolor{mycolor6}{rgb}{0.30100,0.74500,0.93300}%
\definecolor{mycolor7}{rgb}{0.63500,0.07800,0.18400}%
\begin{tikzpicture}
\pgfplotsset{every tick label/.append style={font=\scriptsize}}

\begin{axis}[%
width=0.951\fwidth,
height=\fheight,
at={(0\fwidth,0\fheight)},
scale only axis,
xmin=5,
xmax=45,
xlabel style={font=\footnotesize\color{white!15!black}},
xlabel={UAV-to-UAV distance [m]},
ymin=80,
ymax=110,
ylabel style={font=\footnotesize\color{white!15!black}},
ylabel={Pathloss [dB]},
axis background/.style={fill=white},
axis x line*=bottom,
axis y line*=left,
xmajorgrids,
ymajorgrids,
legend style={legend cell align=left, align=left, draw=white!15!black, font=\scriptsize, at={(0.5,0.98)}, anchor=south},
legend columns=3
]
\addplot [color=mycolor1, dashed]
  table[row sep=crcr]{%
6	85.4226898283147\\
12	92.1317512435563\\
18	96.0563005865078\\
24	98.8408126587978\\
28	100.332857071828\\
32	101.625324731088\\
36	102.765362001749\\
40	103.785160090979\\
};
\addlegendentry{CI fit, $h=6$~m}

\addplot [color=mycolor2, dashdotted]
  table[row sep=crcr]{%
6	85.609557951523\\
9	89.5763945015912\\
12	92.3909098248103\\
15	94.5740175141386\\
18	96.3577463748786\\
21	97.8658676379134\\
24	99.1722616980976\\
27	100.324582924947\\
30	101.355369387426\\
33	102.287829165791\\
36	103.139098248166\\
40	104.169884710645\\
};
\addlegendentry{CI fit, $h=12$~m}

\addplot [color=mycolor3, densely dotted, line width=0.8pt]
  table[row sep=crcr]{%
6	85.7910476854484\\
12	92.642609371789\\
18	96.6505160296761\\
24	99.4941710581296\\
30	101.699881258816\\
36	103.502077716017\\
40	104.54353628727\\
};
\addlegendentry{CI fit, $h=15$~m}

\addplot [color=mycolor4]
  table[row sep=crcr]{%
6	85.5996542949231\\
9	89.5642497031066\\
12	92.3771749108529\\
15	94.5590492107989\\
18	96.3417703190363\\
21	97.8490395394706\\
24	99.1546955267826\\
27	100.30636572722\\
30	101.336569826729\\
33	102.268502793707\\
36	103.119290934966\\
39	103.9019401567\\
42	104.6265601554\\
};
\addlegendentry{CI fit, all heights}


\end{axis}

\begin{axis}[%
width=0.951\fwidth,
height=\fheight,
at={(0\fwidth,0\fheight)},
scale only axis,
xmin=5,
xmax=45,
xlabel style={font=\footnotesize\color{white!15!black}},
ymin=80,
ymax=110,
ylabel style={font=\footnotesize\color{white!15!black}},
hide x axis,
hide y axis,
legend style={legend cell align=left, align=left, draw=white!15!black, font=\scriptsize, at={(0.99,0.01)}, anchor=south east},
legend columns=1
]
\addplot[only marks, mark=asterisk, mark options={}, mark size=1.5000pt, draw=mycolor1] table[row sep=crcr]{%
x	y\\
6	83.1280544048846\\
12	88.9876700318486\\
18	93.4918541481491\\
24	102.287628326341\\
30	100.058029873338\\
36	105.505665325063\\
40	107.934794437175\\
};
\addlegendentry{Meas., $h = 6$ m}

\addplot[only marks, mark=+, mark options={}, mark size=1.5000pt, draw=mycolor6] table[row sep=crcr]{%
x	y\\
6	86.0397539475613\\
9	91.5460418808438\\
12	94.4062366913422\\
15	97.9879394678565\\
18	95.7170813033622\\
21	95.3777219945156\\
24	99.02263962\\
27	99.2763817738388\\
30	102.051493191169\\
33	100.437992682641\\
36	102.844470979747\\
40	103.9426507\\
};
\addlegendentry{Meas., $h = 12$ m}

\addplot[only marks, mark=o, mark options={}, mark size=1.5000pt, draw=mycolor7] table[row sep=crcr]{%
x	y\\
6	85.2846006761252\\
12	91.9142447575216\\
18	97.7809256435052\\
24	97.5739435439125\\
28	100.3026762\\
32	101.3044984\\
36	101.9418107\\
40	104.8675908\\
};
\addlegendentry{Meas., $h = 15$ m}

\end{axis}
\end{tikzpicture}%
    \caption{Comparison of different CI fits considering measurements at different heights, and a CI fit that combines all the measurements.}
    \label{fig:ci_height}
\end{figure}

As the deployment height of \gls{uav} networks is subject to application scenario, regulations, and performance requirements~\cite{amorim2017radio}, it is important to characterize the channel behavior for a wide range of deployment heights. Therefore, in Fig.~\ref{fig:ci_height} we investigate the impact of the height on the parameters of a \gls{ci} fit. For this, we consider separate \gls{ci} fits for the three values of the \glspl{uav} height at which the measurements were collected, i.e., $h=6$, $12$, and \meter{15}, and compare them with a \gls{ci} fit that does not distinguish between different heights. As can be seen, the four curves in Fig.~\ref{fig:ci_height} share the same trend, with differences of less than 1 dB between the curves for $h=6$ m and $h=15$ m in the worst case. The path loss exponent $n_{CI}$ and the standard deviation are shown in Table~\ref{tab:ci_height}. The value of $n_{CI}$ is very similar for the four fits, showing that the height does not a significant impact once the \glspl{uav} are in flight. The standard deviation $\sigma$ shows a higher variability, also with respect to the \gls{fi} fit previously described, but this can be traced back to the fewer measurements considered for the fit at each single height value.

\subsection{Comparison with Free Space and 3GPP Path Loss Models}

\begin{figure}[t]
    \centering
    \setlength\fheight{.4\columnwidth}
    \setlength\fwidth{.9\columnwidth}
%
%
\definecolor{mycolor1}{rgb}{0.00000,0.44700,0.74100}%
\definecolor{mycolor2}{rgb}{0.85000,0.32500,0.09800}%
\definecolor{mycolor3}{rgb}{0.92900,0.69400,0.12500}%
\definecolor{mycolor4}{rgb}{0.49400,0.18400,0.55600}%
\definecolor{mycolor5}{rgb}{0.46600,0.67400,0.18800}%
\definecolor{mycolor6}{rgb}{0.30100,0.74500,0.93300}%
\begin{tikzpicture}
\pgfplotsset{every tick label/.append style={font=\scriptsize}}
\pgfplotsset{every axis label/.append style={font=\footnotesize}}

\begin{axis}[%
width=0.951\fwidth,
height=\fheight,
at={(0\fwidth,0\fheight)},
scale only axis,
xmin=5,
xmax=40,
xlabel style={font=\footnotesize\color{white!15!black}},
xlabel={UAV-to-UAV distance [m]},
ymin=80,
ymax=105,
ylabel style={font=\footnotesize\color{white!15!black}},
ylabel={Pathloss [dB]},
axis background/.style={fill=white},
axis x line*=bottom,
axis y line*=left,
xmajorgrids,
ymajorgrids,
legend style={legend cell align=left, align=left, draw=white!15!black, font=\scriptsize, at={(0.5, 0.95)}, anchor=south},
legend columns = 3
]
\addplot [color=mycolor1,dashdotdotted, mark=o, mark options={solid, mycolor1}, mark size=1pt]
  table[row sep=crcr]{%
6	85.5996542949231\\
9	89.5642497031066\\
12	92.3771749108529\\
15	94.5590492107989\\
18	96.3417703190363\\
21	97.8490395394706\\
24	99.1546955267826\\
27	100.30636572722\\
30	101.336569826729\\
33	102.268502793707\\
36	103.119290934966\\
39	103.9019401567\\
42	104.6265601554\\
};
\addlegendentry{UAV CI fit}

\addplot [color=mycolor2, dashed]
  table[row sep=crcr]{%
6	84.4622599079195\\
8.42857142857143	87.5978793397629\\
10.8571428571429	89.9430348188888\\
13.2857142857143	91.8200535863391\\
15.7142857142857	93.3871203407435\\
18.1428571428571	94.7337143782101\\
20.5714285714286	95.9154108578496\\
23	96.9691042062325\\
25.4285714285714	97.9205431437647\\
27.8571428571429	98.7884122196051\\
30.2857142857143	99.5867007462102\\
32.7142857142857	100.326155081558\\
35.1428571428571	101.015209487306\\
37.5714285714286	101.660605242134\\
40	102.26781546775\\
};
\addlegendentry{3GPP UMi model}

\addplot [color=mycolor3, densely dotted, line width=0.8pt]
  table[row sep=crcr]{%
6	80.8404111583032\\
8.42857142857143	84.1236333113908\\
10.8571428571429	86.5787503711553\\
13.2857142857143	88.5434384948788\\
15.7142857142857	90.1834149858875\\
18.1428571428571	91.5924200591518\\
20.5714285714286	92.8286753099306\\
23	93.9308320422501\\
25.4285714285714	94.9258651060593\\
27.8571428571429	95.8333487909533\\
30.2857142857143	96.6679385671247\\
32.7142857142857	97.440892523884\\
35.1428571428571	98.1610465543951\\
37.5714285714286	98.8354629506097\\
40	99.4698754590782\\
};
\addlegendentry{3GPP UMa model}

\addplot [color=mycolor4, dashdotted]
  table[row sep=crcr]{%
6	83.4052173957303\\
8.42857142857143	86.4671704956753\\
10.8571428571429	88.7583115591957\\
13.2857142857143	90.5929543715165\\
15.7142857142857	92.1253509630597\\
18.1428571428571	93.4427560634227\\
20.5714285714286	94.5993631641892\\
23	95.6311494506308\\
25.4285714285714	96.5632234979624\\
27.8571428571429	97.4138053342286\\
30.2857142857143	98.1965364801919\\
32.7142857142857	98.9218961260148\\
35.1428571428571	99.5981088471266\\
37.5714285714286	100.231748325642\\
40	100.82815161466\\
};
\addlegendentry{3GPP RMa model}

\addplot [color=mycolor6, dashed]
  table[row sep=crcr]{%
6	81.5831002815\\
8.42857142857143	84.1725896447397\\
10.8571428571429	86.1108872755026\\
13.2857142857143	87.6635294247423\\
15.7142857142857	88.9608301537108\\
18.1428571428571	90.0765033587258\\
20.5714285714286	91.0563323851499\\
23	91.9307112129674\\
25.4285714285714	92.7208518832745\\
27.8571428571429	93.4421469056165\\
30.2857142857143	94.1061208088266\\
32.7142857142857	94.7216265449536\\
35.1428571428571	95.2956123390763\\
37.5714285714286	95.8336317207749\\
40	96.3401934998368\\
};
\addlegendentry{3GPP InOo model}

\addplot [color=mycolor5, solid, mark=x, mark options={solid, mycolor5}, mark size=1pt]
  table[row sep=crcr]{%
6	83.6430426625529\\
9	87.1648678436666\\
12	89.6636425758326\\
15	91.6018428359937\\
18	93.1854677569462\\
21	94.5244035495585\\
24	95.6842424891122\\
27	96.7072929380598\\
30	97.6224427492733\\
33	98.4502964524378\\
36	99.2060676702258\\
39	99.9013097954101\\
42	100.545003462838\\
};
\addlegendentry{Free space pathloss}

\end{axis}
\end{tikzpicture}%
    \setlength\abovecaptionskip{-.1cm}
    \setlength\belowcaptionskip{-.3cm}
    \caption{Comparison of CI fit, 3GPP and free space path loss channel models.}
    \label{fig:ci_3gpp}
\end{figure}
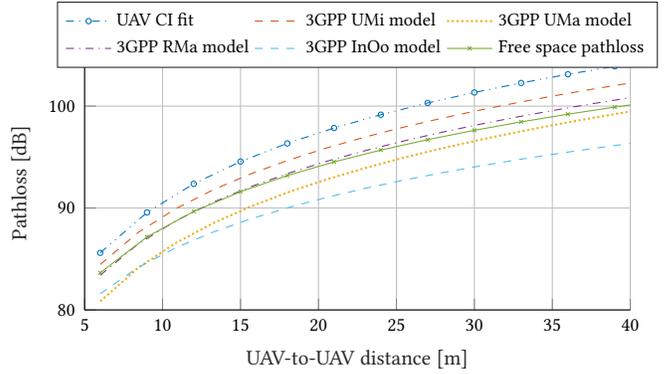

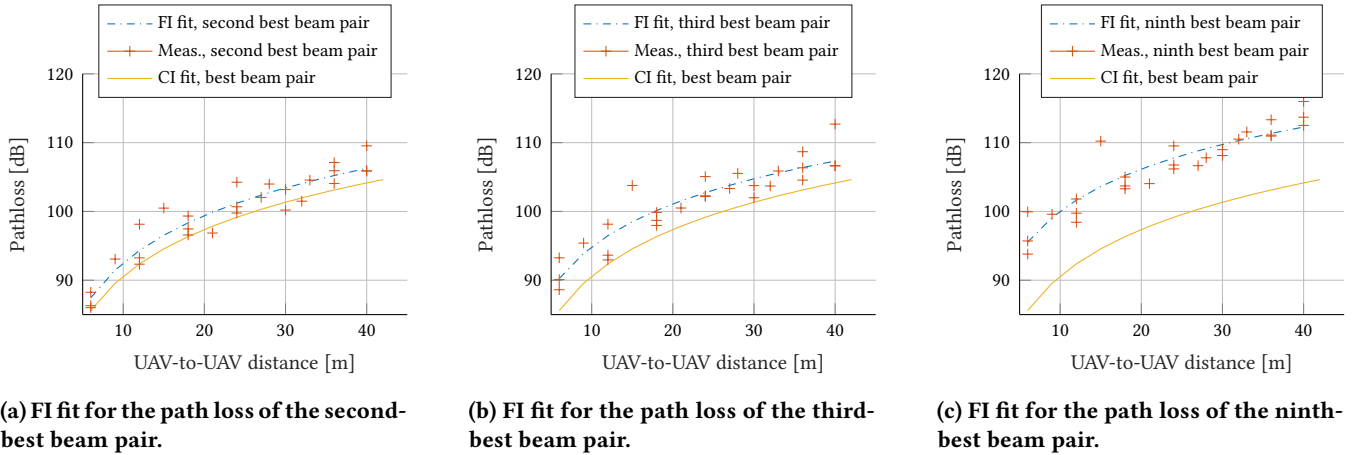
\begin{figure*}[b]
\begin{subfigure}[t]{0.3\textwidth}
    \centering
    \setlength\fheight{.6\columnwidth}
    \setlength\fwidth{.85\columnwidth}
%
%
\definecolor{mycolor1}{rgb}{0.00000,0.44700,0.74100}%
\definecolor{mycolor2}{rgb}{0.85000,0.32500,0.09800}%
\definecolor{mycolor3}{rgb}{0.92900,0.69400,0.12500}%
\begin{tikzpicture}
\pgfplotsset{every tick label/.append style={font=\scriptsize}}

\begin{axis}[%
width=0.951\fwidth,
height=\fheight,
at={(0\fwidth,0\fheight)},
scale only axis,
xmin=5,
xmax=45,
xlabel style={font=\footnotesize\color{white!15!black}},
xlabel={UAV-to-UAV distance [m]},
ymin=85,
ymax=120,
ylabel style={font=\footnotesize\color{white!15!black}},
ylabel={Pathloss [dB]},
axis background/.style={fill=white},
axis x line*=bottom,
axis y line*=left,
xmajorgrids,
ymajorgrids,
legend style={legend cell align=left, align=left, draw=white!15!black, font=\scriptsize, at={(0.5,0.9)},anchor=south}
]
\addplot [color=mycolor1, dashdotted]
  table[row sep=crcr]{%
6	87.4511117039354\\
9	91.4721616275038\\
12	94.3251419513634\\
15	96.538085413116\\
18	98.3461918749317\\
21	99.8749241059972\\
24	101.199172198791\\
27	102.3672417985\\
28	102.727904429857\\
30	103.412115660544\\
32	104.052152522651\\
33	104.357319041929\\
36	105.22022212236\\
40	106.265095984404\\
};
\addlegendentry{\gls{fi} fit, second best beam pair}

\addplot [color=mycolor2, draw=none, mark=+, mark options={solid, mycolor2}]
  table[row sep=crcr]{%
6	86.3015374201197\\
12	93.2551855681366\\
18	99.3315044136944\\
24	99.7704273623075\\
28	103.9830201\\
32	101.4869628\\
36	104.0580145\\
40	105.8737703\\
6	88.2490421650244\\
9	93.0762607644929\\
12	98.138042019258\\
15	100.482864025728\\
18	97.4578698788644\\
21	96.8624717415111\\
24	100.67997\\
27	102.036977970232\\
30	103.183984269824\\
33	104.556094648493\\
36	105.924311991496\\
40	105.9047062\\
6	86.0134438408899\\
12	92.3303255994854\\
18	96.5789590305291\\
24	104.246271661375\\
30	100.199907828174\\
36	107.113872160542\\
40	109.539029507821\\
};
\addlegendentry{Meas., second best beam pair}

\addplot [color=mycolor3]
  table[row sep=crcr]{%
6	85.5996542949231\\
9	89.5642497031066\\
12	92.3771749108529\\
15	94.5590492107989\\
18	96.3417703190363\\
21	97.8490395394706\\
24	99.1546955267826\\
27	100.30636572722\\
30	101.336569826729\\
33	102.268502793707\\
36	103.119290934966\\
39	103.9019401567\\
42	104.6265601554\\
};
\addlegendentry{CI fit, best beam pair}

\end{axis}
\end{tikzpicture}%
    \caption{\gls{fi} fit for the path loss of the second-best beam pair.}
    \label{fig:abg_second_best}
\end{subfigure}\hfill
\begin{subfigure}[t]{0.3\textwidth}
    \centering
    \setlength\fheight{.6\columnwidth}
    \setlength\fwidth{.85\columnwidth}
%
%
\definecolor{mycolor1}{rgb}{0.00000,0.44700,0.74100}%
\definecolor{mycolor2}{rgb}{0.85000,0.32500,0.09800}%
\definecolor{mycolor3}{rgb}{0.92900,0.69400,0.12500}%
\begin{tikzpicture}
\pgfplotsset{every tick label/.append style={font=\scriptsize}}

\begin{axis}[%
width=0.951\fwidth,
height=\fheight,
at={(0\fwidth,0\fheight)},
scale only axis,
xmin=5,
xmax=45,
xlabel style={font=\footnotesize\color{white!15!black}},
xlabel={UAV-to-UAV distance [m]},
ymin=85,
ymax=120,
ylabel style={font=\footnotesize\color{white!15!black}},
ylabel={Pathloss [dB]},
axis background/.style={fill=white},
axis x line*=bottom,
axis y line*=left,
xmajorgrids,
ymajorgrids,
legend style={legend cell align=left, align=left, draw=white!15!black, font=\scriptsize, at={(0.5,0.9)},anchor=south}
]
\addplot [color=mycolor1, dashdotted]
  table[row sep=crcr]{%
6	90.2407184037906\\
9	93.8934980199714\\
12	96.4851864024701\\
15	98.4954560888701\\
18	100.137966018651\\
21	101.526688376835\\
24	102.72965440115\\
27	103.790745634832\\
28	104.118376759334\\
30	104.73992408755\\
32	105.321342783648\\
33	105.598560441312\\
36	106.38243401733\\
40	107.331612470048\\
};
\addlegendentry{\gls{fi} fit, third best beam pair}

\addplot [color=mycolor2, draw=none, mark=+, mark options={solid, mycolor2}]
  table[row sep=crcr]{%
6	88.6187007313674\\
12	93.6395038520976\\
18	99.8545666570619\\
24	102.172497674209\\
28	105.5261705\\
32	103.700595\\
36	104.5529496\\
40	106.6530452\\
6	90.096876881913\\
9	95.4001761075336\\
12	98.1443509943084\\
15	103.773899047793\\
18	98.7009489391402\\
21	100.496476340969\\
24	102.2699562\\
27	103.336736236676\\
30	103.757900697047\\
33	105.878038813043\\
36	106.346843203466\\
40	106.6214385\\
6	93.2544915030647\\
12	92.9563499739309\\
18	97.9575259512735\\
24	105.06956519238\\
30	101.989825229993\\
36	108.681629057524\\
40	112.696173335429\\
};
\addlegendentry{Meas., third best beam pair}

\addplot [color=mycolor3]
  table[row sep=crcr]{%
6	85.5996542949231\\
9	89.5642497031066\\
12	92.3771749108529\\
15	94.5590492107989\\
18	96.3417703190363\\
21	97.8490395394706\\
24	99.1546955267826\\
27	100.30636572722\\
30	101.336569826729\\
33	102.268502793707\\
36	103.119290934966\\
39	103.9019401567\\
42	104.6265601554\\
};
\addlegendentry{CI fit, best beam pair}

\end{axis}
\end{tikzpicture}%
    \caption{\gls{fi} fit for the path loss of the third-best beam pair.}
    \label{fig:abg_third_best}
\end{subfigure}\hfill
\begin{subfigure}[t]{0.3\textwidth}
    \centering
    \setlength\fheight{.6\columnwidth}
    \setlength\fwidth{.85\columnwidth}
%
%
\definecolor{mycolor1}{rgb}{0.00000,0.44700,0.74100}%
\definecolor{mycolor2}{rgb}{0.85000,0.32500,0.09800}%
\definecolor{mycolor3}{rgb}{0.92900,0.69400,0.12500}%
\begin{tikzpicture}
\pgfplotsset{every tick label/.append style={font=\scriptsize}}

\begin{axis}[%
width=0.951\fwidth,
height=\fheight,
at={(0\fwidth,0\fheight)},
scale only axis,
xmin=5,
xmax=45,
xlabel style={font=\footnotesize\color{white!15!black}},
xlabel={UAV-to-UAV distance [m]},
ymin=85,
ymax=120,
ylabel style={font=\footnotesize\color{white!15!black}},
ylabel={Pathloss [dB]},
axis background/.style={fill=white},
axis x line*=bottom,
axis y line*=left,
xmajorgrids,
ymajorgrids,
legend style={legend cell align=left, align=left, draw=white!15!black, font=\scriptsize, at={(0.5,0.9)},anchor=south}
]
\addplot [color=mycolor1, dashdotted]
  table[row sep=crcr]{%
6	95.5344195449045\\
9	99.1115561494945\\
12	101.649574961158\\
15	103.618215294252\\
18	105.226711565748\\
21	106.586675785617\\
24	107.764730377411\\
27	108.803848170338\\
28	109.12469459728\\
30	109.733370710506\\
32	110.302749189074\\
33	110.574226128519\\
36	111.341866982001\\
40	112.271389522169\\
};
\addlegendentry{\gls{fi} fit, ninth best beam pair}

\addplot [color=mycolor2, draw=none, mark=+, mark options={solid, mycolor2}]
  table[row sep=crcr]{%
6	95.7068997443409\\
12	99.7474504957271\\
18	103.293464964512\\
24	106.752622791732\\
28	107.7979046\\
32	110.5072082\\
36	110.9541186\\
40	112.5107709\\
6	93.7977607672313\\
9	99.5887946718305\\
12	101.811999082731\\
15	110.22094513578\\
18	105.004772387784\\
21	104.047078800476\\
24	106.1824511\\
27	106.649216363144\\
30	108.126697042084\\
33	111.558753596177\\
36	111.089970093246\\
40	113.6975474\\
6	99.9460974846857\\
12	98.4266376122825\\
18	103.708396884045\\
24	109.527259589961\\
30	108.984362800454\\
36	113.331447770336\\
40	115.984156717201\\
};
\addlegendentry{Meas., ninth best beam pair}

\addplot [color=mycolor3]
  table[row sep=crcr]{%
6	85.5996542949231\\
9	89.5642497031066\\
12	92.3771749108529\\
15	94.5590492107989\\
18	96.3417703190363\\
21	97.8490395394706\\
24	99.1546955267826\\
27	100.30636572722\\
30	101.336569826729\\
33	102.268502793707\\
36	103.119290934966\\
39	103.9019401567\\
42	104.6265601554\\
};
\addlegendentry{CI fit, best beam pair}

\end{axis}
\end{tikzpicture}%
    \caption{\gls{fi} fit for the path loss of the ninth-best beam pair.}
    \label{fig:abg_ninth_best}
\end{subfigure}\hfill

    \caption{Measurements vs. \gls{fi} fits for the path loss of additional beam pairs.}
    \label{fig:other_beams}
\end{figure*}

\begin{table*}[b]
    \centering
    \footnotesize
    \begin{tabular}{llllllllll}
    \toprule
        & Best beam pair & 2nd best & 3rd best & 4th best & 5th best & 6th best & 7th best & 8th best & 9th best\\\midrule
    PLE $n$ & 2.25 & 2.28 & 2.07 & 1.96 & 2.01 & 1.93 & 1.99 & 2.02 & 2.03 \\
    Intercept [dB] & 68.08 & 69.68 & 74.10 & 76.79 & 77.26 & 79.31 & 79.35 & 79.52 & 79.73 \\ 
    $\sigma$ [dB] & 3.56 & 3.78 &   4.85 &  4.61 &  4.01 &  5.76 &  5.80 &  5.38 &  4.82 \\
    Displacement $\Delta$ [deg] & 0 &  1.87 & 2.59 & 2.70 & 3.47 & 3.42 & 4.20 & 3.89 & 4.20 \\
    \bottomrule
    \end{tabular}
    \caption{Parameters for the fit for different best beam pairs.}
    \label{tab:abg_beams}
\end{table*}

As discussed in Sec.~\ref{sec:intro}, aerial communications are affected by a different channel than terrestrial networks, where both endpoints are on the ground and present less erratic mobility patterns. An aerial link is indeed characterized by a particularly strong \gls{los} link, while reflections provide limited or no contribution, as scatterers are at a larger distance than in cellular or indoor scenarios. Furthermore, the micro-mobility of the \glspl{uav}, given by the fluctuations caused by the \gls{uav} control loop and/or wind conditions, may degrade the link quality~\cite{DabiriTWC}.

Figure~\ref{fig:ci_3gpp} shows that the measurement-based fit we propose in this paper reflects these phenomena into a higher path loss exponent with respect to free space path loss and of \gls{3gpp} channel models. We consider the equations for \gls{los} propagation loss from the \gls{3gpp} channel model for frequencies between 0.5 and 100 GHz~\cite{3gpp.38.901}, with an additional distance-dependent oxygen absorption loss factor for the 60 GHz band. We compare the path loss of different \gls{3gpp} scenarios, i.e., for \gls{rma}, \gls{uma}, \gls{umi} and \gls{inoo} deployments. The free space path loss follows Friis' law~\cite{rappaport2017overview}. The path loss curves in Fig.~\ref{fig:ci_3gpp} confirm that the propagation of 60 GHz signals experiences a higher loss in an aerial link, as it has a path loss exponent $n_{CI} = 2.25$ which is larger than the worse \gls{los} exponent in \gls{3gpp} scenarios (i.e., 2.1 for \gls{umi}). This confirms that \textit{the analysis and simulation of aerial networks cannot be based on channel models developed for terrestrial applications}, and motivates the development of the experimental channel model proposed in this paper.

\subsection{Path Loss Curves with Misalignment}

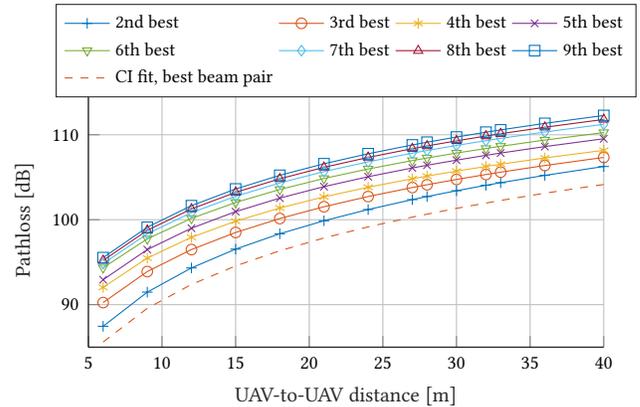
\begin{figure}[t]
    \centering
    \setlength\fheight{.4\columnwidth}
    \setlength\fwidth{.85\columnwidth}
%
%
\definecolor{mycolor1}{rgb}{0.00000,0.44700,0.74100}%
\definecolor{mycolor2}{rgb}{0.85000,0.32500,0.09800}%
\definecolor{mycolor3}{rgb}{0.92900,0.69400,0.12500}%
\definecolor{mycolor4}{rgb}{0.49400,0.18400,0.55600}%
\definecolor{mycolor5}{rgb}{0.46600,0.67400,0.18800}%
\definecolor{mycolor6}{rgb}{0.30100,0.74500,0.93300}%
\definecolor{mycolor7}{rgb}{0.63500,0.07800,0.18400}%
\begin{tikzpicture}
\pgfplotsset{every tick label/.append style={font=\scriptsize}}

\begin{axis}[%
width=0.951\fwidth,
height=\fheight,
at={(0\fwidth,0\fheight)},
scale only axis,
xmin=5,
xmax=40,
xlabel style={font=\footnotesize\color{white!15!black}},
xlabel={UAV-to-UAV distance [m]},
ymin=85,
ymax=115,
ylabel style={font=\footnotesize\color{white!15!black}},
ylabel={Pathloss [dB]},
axis background/.style={fill=white},
axis x line*=bottom,
axis y line*=left,
xmajorgrids,
ymajorgrids,
legend style={legend cell align=left, align=left, draw=white!15!black, font=\scriptsize, at={(0.5, 0.98)}, anchor=south},
legend columns=4
]

\addplot [color=mycolor1, mark=+, mark options={solid, mycolor1}]
  table[row sep=crcr]{%
6	87.4511117039354\\
9	91.4721616275038\\
12	94.3251419513634\\
15	96.538085413116\\
18	98.3461918749317\\
21	99.8749241059972\\
24	101.199172198791\\
27	102.3672417985\\
28	102.727904429857\\
30	103.412115660544\\
32	104.052152522651\\
33	104.357319041929\\
36	105.22022212236\\
40	106.265095984404\\
};
\addlegendentry{2nd best}

\addplot [color=mycolor2, mark=o, mark options={solid, mycolor2}]
  table[row sep=crcr]{%
6	90.2407184037906\\
9	93.8934980199714\\
12	96.4851864024701\\
15	98.4954560888701\\
18	100.137966018651\\
21	101.526688376835\\
24	102.72965440115\\
27	103.790745634832\\
28	104.118376759334\\
30	104.73992408755\\
32	105.321342783648\\
33	105.598560441312\\
36	106.38243401733\\
40	107.331612470048\\
};
\addlegendentry{3rd best}

\addplot [color=mycolor3, mark=asterisk, mark options={solid, mycolor3}]
  table[row sep=crcr]{%
6	92.0324691097785\\
9	95.4813262620026\\
12	97.9283293537638\\
15	99.8263724098321\\
18	101.377186505988\\
21	102.688381141509\\
24	103.824189597749\\
27	104.826043658212\\
28	105.13538423327\\
30	105.722232653817\\
32	106.27119268951\\
33	106.532934212903\\
36	107.273046749973\\
40	108.169235745579\\
};
\addlegendentry{4th best}

\addplot [color=mycolor4, mark=x, mark options={solid, mycolor4}]
  table[row sep=crcr]{%
6	92.9339181765057\\
9	96.4812635496612\\
12	98.9981451462384\\
15	100.950390181569\\
18	102.545490519394\\
21	103.894128638727\\
24	105.062372115971\\
27	106.092835892549\\
28	106.411010235304\\
30	107.014617151302\\
32	107.579253712548\\
33	107.848469728459\\
36	108.609717489127\\
40	109.531498747879\\
};
\addlegendentry{5th best}

\addplot [color=mycolor5, mark=triangle, mark options={solid, rotate=180, mycolor5}]
  table[row sep=crcr]{%
6	94.3409574216058\\
9	97.7432470744755\\
12	100.157209999635\\
15	102.029625111463\\
18	103.559499652505\\
21	104.852990146437\\
24	105.973462577664\\
27	106.961789305374\\
28	107.266953071596\\
30	107.845877689492\\
32	108.387425502823\\
33	108.645632913991\\
36	109.375752230534\\
40	110.259840614652\\
};
\addlegendentry{6th best}

\addplot [color=mycolor6, mark=diamond, mark options={solid, mycolor6}]
  table[row sep=crcr]{%
6	94.8350194062806\\
9	98.3399790307704\\
12	100.826787460077\\
15	102.755705934643\\
18	104.331747084567\\
21	105.664270890137\\
24	106.818555513873\\
27	107.836706709057\\
28	108.151079319443\\
30	108.747473988439\\
32	109.30536394318\\
33	109.571363209328\\
36	110.323515138363\\
40	111.234282417746\\
};
\addlegendentry{7th best}

\addplot [color=mycolor7, mark=triangle, mark options={solid, mycolor7}]
  table[row sep=crcr]{%
6	95.2026309989766\\
9	98.7513860772727\\
12	101.269267875565\\
15	103.222288727618\\
18	104.818022953861\\
21	106.167197018214\\
24	107.335904752153\\
27	108.366778032157\\
28	108.685078816506\\
30	109.288925604206\\
32	109.853786550445\\
33	110.123109552048\\
36	110.884659830449\\
40	111.806807402498\\
};
\addlegendentry{8th best}

\addplot [color=mycolor1, mark=square, mark options={solid, mycolor1}]
  table[row sep=crcr]{%
6	95.5344195449045\\
9	99.1115561494945\\
12	101.649574961158\\
15	103.618215294252\\
18	105.226711565748\\
21	106.586675785617\\
24	107.764730377411\\
27	108.803848170338\\
28	109.12469459728\\
30	109.733370710506\\
32	110.302749189074\\
33	110.574226128519\\
36	111.341866982001\\
40	112.271389522169\\
};
\addlegendentry{9th best}

\addplot [color=mycolor2, dashed]
  table[row sep=crcr]{%
6	85.5996542949231\\
9	89.5642497031066\\
12	92.3771749108529\\
15	94.5590492107989\\
18	96.3417703190363\\
21	97.8490395394706\\
24	99.1546955267826\\
27	100.30636572722\\
30	101.336569826729\\
33	102.268502793707\\
36	103.119290934966\\
39	103.9019401567\\
42	104.6265601554\\
};
\addlegendentry{CI fit, best beam pair}

\end{axis}
\end{tikzpicture}%
    \caption{\gls{fi} fit for the path loss of first 9 beam pairs.}
    \label{fig:abg_beams}
\end{figure}

\gls{mmwave} links will rely on beamforming techniques to increase the link budget and compensate for the increased path loss. In this sense, a proper beam alignment makes it possible to select the transmit and receive beams that yield the highest beamforming gain. Fast and prompt beam tracking, however, can be challenging in aerial links, where the nodes are highly mobile~\cite{BertizzoloMmnets19}. Therefore, we present in the following paragraph the path loss curves that fit measurements for beam pairs that are not perfectly aligned, following a similar approach discussed in~\cite{preibe2012affection} for terahertz links. Notice that these results do not represent the actual path loss (which is given only by the data for the best beam pair), but are a practical way to model the loss in beamforming gain due to misalignment.

For this analysis, we exploit the beam scanning capabilities of the Terragraph channel sounders~\cite{terragraphindoor}. For each distance and height at which measurements are taken, the beam pair that yields the lowest path loss is considered for the \gls{ci} fit previously described. We then analyze the remaining beam pairs, and, for each distance-height point, select the second-through-ninth best beam pairs in terms of path loss. Finally, we apply an \gls{fi} fit for each of these, by aggregating the data for different heights. For this analysis, the \gls{fi} fit is preferred over the \gls{ci} fit, as its intercept term $PL_{FI}$ allows the modeling of the additional loss (with respect to the free space path loss) introduced by beam misalignment. 

As an example, Fig.~\ref{fig:other_beams} reports the measurements and fitted path loss curves for the second, third and ninth-best beam pairs, and the path loss curve for the best beam pair. We observe that the distance between measured points and the best beam pair fit line progressively increases, as the misalignment between the non-optimal beam pairs decreases the beamforming gain.
Figure~\ref{fig:abg_beams}, instead, directly compares path loss curves, which exhibit the same trend, as the physical path loss between the two nodes does not change, but have an increasing loss factor that models the reduction in beamforming gain. 

To further underline this, we list in Table~\ref{tab:abg_beams} the path loss exponents, which are comparable for the different beam pairs, and the intercept values, which increase by up to $11\:\mathrm{dB}$ from the best to the ninth-best beam pair. We also provide a displacement metric, to model the angular error that can lead to such loss in beamforming gain. Notably, for the $i$-th-best beam pair, $i \ge 2$, we have $\Delta_i = \mathop{\mathbb{E}}\left[|\theta_{tx, best} - \theta_{tx, i}| + |\theta_{rx, best} - \theta_{rx, i}|\right]$, with $\mathop{\mathbb{E}}$ the average operator, and $\theta_{j, best}$ and $\theta_{j, i}$ the angles of the best beam, and of the beam associated to the $i$-th beam pair, for $j \in \{tx, rx\}$, respectively. As discussed in Sec.~\ref{sec:meas}, the beams of the Terragraph sounders are spaced by $1.4^\circ$. Therefore, $\Delta = 1.8667^\circ$, for the second beam pair, suggests that, on average, the beamforming gain loss is due to a misalignment of 1.33 beam pairs, either at the receiver or the transmitter. We notice that $\Delta$ saturates after the fifth-best beam pair, and that the intercept $PL_{FI}$ has a limited difference, showing that the misalignment is likely due to the selection of equivalent beam pairs.

\section{Conclusions}
\label{sec:conclusions}

In this paper, we described an extensive measurement campaign for the characterization of the \gls{a2a} path loss at \freqGhz{60}, using the \terragraph radios configured as channel sounders. The data analysis led us to conclude that the propagation loss of an aerial link, for heights between $6$ and \meter{15}, is well represented by the equation:
\begin{equation}
\setlength{\abovedisplayskip}{1.5pt}
\setlength{\belowdisplayskip}{1.5pt}
    PL_{CI}(d) = 68.08 + 22.5 \log_{10}(d) + \xi_{\sigma, CI},
\end{equation}
with $\xi_{\sigma, CI}$ a Gaussian random variable with standard deviation $\sigma = 3.56$ dB. Moreover, we compared this fit with established propagation models for \glspl{mmwave}, and analyzed the impact of a sub-optimal beam selection on the link performance. For this case, the combined path loss and gain reduction can be computed using Eq.~\eqref{eq:fi} with the parameters from Table~\ref{tab:abg_beams}.

As part of our future work, we will extend the validation of the path loss curves by performing measurements in different scenarios and with different sounders, and analyze the impact of beam misalignment with different beamforming configurations, as well as of the Doppler effect in highly mobile conditions. Moreover, we will characterize \gls{a2g} \gls{mmwave} channels and extend the height range for the  \gls{a2a} measurements. 

\section*{Acknowledgements}
This work was supported in part by the US National Science Foundation
under Grant CNS-1618727 and in part by the US Office of Naval Research
under Grants N00014-19-1-2409 and N00014-20-1-2132.

\bibliographystyle{ACM-Reference-Format.bst}
\bibliography{bibl.bib}

\end{document}